\documentclass[fleqn,11pt]{article}%
\usepackage{amsmath}
\usepackage{float}
\usepackage{theorem}
\usepackage{graphicx}%
\usepackage{amsfonts}%
\usepackage{amssymb}
\theorembodyfont{\upshape}
\newtheorem{theorem}{Theorem}

\newtheorem{definition}[theorem]{Definition}
\newtheorem{example}[theorem]{Example}

\newtheorem{remark}[theorem]{Remark}

\advance\topmargin by -0.75truein
\advance\textheight by 1.25truein
\begin{document}

\begin{center}
$\int${\LARGE On ``\textit{Exponential Lower Bounds for Polytopes in
Combinatorial Optimization''}} {\LARGE by Fiorini \textit{et al}.
(2015)}$^{^{\ast}}${\LARGE : }\newline {\LARGE A Refutation For Models With
Disjoint Sets of Descriptive Variables }{\Large \medskip\medskip}

Moustapha Diaby

OPIM Department; University of Connecticut; Storrs, CT 06268\\[0pt]%
moustapha.diaby@business.uconn.edu\medskip{\Large \medskip}

Mark H. Karwan

Department of Industrial and Systems Engineering; SUNY at Buffalo; Amherst, NY
14260\\[0pt]mkarwan@buffalo.edu\medskip{\Large \medskip}

Lei Sun

Praxair, Inc.; Tonawanda, NY 14150\\[0pt]leisun@buffalo.edu\medskip
\end{center}

\textsl{Abstract.}{\small \ We provide a numerical refutation of the
developments of Fiorini \textit{et al.} (2015)}$^{{\small \ast}}${\small \ for
models with disjoint sets of descriptive variables. We also provide an insight
into the meaning of the existence of a one-to-one linear map between solutions
of such models.\bigskip}

\textsl{Keywords:}\textbf{\ }{\small Linear Programming; Combinatorial
Optimization; Traveling Salesman Problem; TSP; Computational
Complexity.\medskip}

{\small *: Fiorini, S., S. Massar, S. Pokutta, H.R. Tiwary, and R. de Wolf
(2015). Exponential Lower Bounds for Polytopes in Combinatorial Optimization.
\textit{Journal of the ACM }62:2, Article No. 17.}

\section{Introduction\label{Introduction_Section}}

\noindent\textit{Extended formulations }(EFs) have been the dominant theory
which has been used in deciding on the validity of proposed LP models for hard
combinatorial problems. All of the developments are predicated on the model
being evaluated \textit{projecting} to the ``natural'' polytope of the
specific problem at hand. We show in Diaby and Karwan (2016a and 2016b,
respectively) that the developments of Fiorini \textit{et al}. (2011; 2012)
are not applicable in relating polytopes which are described in terms of
disjoint sets of variables. In the journal version of their papers, Fiorini
\textit{et al.} (Fiorini \textit{et al.} (2015)) have added the stipulation
that the polytopes they consider be of dimensions greater than zero (e.g., see
p. 17:9, last sentence of ``Lemma 2''; p. 17:17, first sentence of ``Theorem
13''). The objective of this technical note is to show that the
counter-examples provided in Diaby and Karwan (2016a and 2016b, respectively)
for the Fiorini \textit{et al}. (2011; 2012) developments remain valid for the
Fiorini \textit{et al}. (2015) work. We do this by providing a counter-example
using polytopes of dimensions greater than zero which similarly refute the
Fiorini \textit{et al}. (2015) developments. For convenience, we start with a
recall of the standard definition of an ``\textit{extended formulation}'' as
well as those of the alternate definitions used in Fiorini \textit{et al}.
(2015). Then, we discuss our numerical example. Finally, we provide some
insight into the meaning of the existence of a one-to-one correspondence
between solutions of models when the models have disjoint sets of descriptive variables.

\section{Background Definitions}

\begin{definition}
[``Standard EF Definition'']\label{EF_Dfn_Std}An \textit{extended formulation}
for a polytope $X$ $\subseteq$ $\mathbb{R}^{p}$ is a polyhedron $U=\left\{
\dbinom{w}{x}\in\mathbb{R}^{p+q}:Gx+Hw\leq g\right\}  $ the projection,
$\varphi_{x}(U):=\left\{  x\in\mathbb{R}^{p}:\right.  \left(  \exists
w\in\mathbb{R}^{q}:\right.  $ $\left.  \left.  \dbinom{w}{x}\in U\right)
\right\}  ,$ of which onto $x$-space is equal to $X$ (where $G$ $\in
\mathbb{R}^{m\times p},$ $H\in\mathbb{R}^{m\times q},$ and $g\in\mathbb{R}%
^{m}$) (Yannakakis (1991)).
\end{definition}

\begin{definition}
[``Fiorini \textit{et al.} Definition \#1'']\label{EF_Dfn_A1}A polyhedron $U$
$=$ $\left\{  \dbinom{w}{x}\in\mathbb{R}^{p+q}:Gx+Hw\right.  $ $\left.  \leq
g\right\}  $ is an \textit{extended formulation} of a polytope $X$ $\subseteq$
$\mathbb{R}^{p}$ if there exists a linear map $\pi$ $:$ $\mathbb{R}^{p+q}$
$\longrightarrow$ $\mathbb{R}^{p}$ such that $X$ is the image of $U$ under
$\pi$ (i.e., $X=\pi(U)$; where $G\in\mathbb{R}^{m\times p}$, $H\in
\mathbb{R}^{m\times q},$ and $g\in\mathbb{R}^{m}$) (see Fiorini \textit{et al}
(2015; p. 17:3, lines 20-21; p. 17:9, lines 22-23)).
\end{definition}

\begin{definition}
[``Fiorini \textit{et al.} Definition \#2'']\label{EF_Dfn_A2}An
\textit{extended formulation} of a polytope $X$ $\subseteq$ $\mathbb{R}^{p}$
is a linear system $U$ $=$ $\left\{  \dbinom{w}{x}\in\mathbb{R}^{p+q}%
:Gx+Hw\leq g\right\}  $ such that $x\in X$ if and only if there exists
$w\in\mathbb{R}^{q}$ such that $\dbinom{w}{x}\in U.$ (In other words, $U$ is
an EF of $X$ if $\left(  x\in X\Longleftrightarrow\left(  \exists
w\in\mathbb{R}^{q}:\dbinom{w}{x}\in U\right)  \right)  $ (where $G$
$\in\mathbb{R}^{m\times p},$ $H\in\mathbb{R}^{m\times q},$ and $g\in
\mathbb{R}^{m}$) (see Fiorini \textit{et al}. (2015; p. 17:2, last paragraph;
p. 17:9, line 20-21)).$\medskip$
\end{definition}

\section{Numerical refutation of Fiorini \textit{et al.} (2015)}

Our numerical counter-example will now be discussed.\medskip

\begin{example}
\label{No_EF-Relation_Example}: Let $\mathbf{x}\in\mathbb{R}^{3}$ and
$\mathbf{w\in}\mathbb{R}$ be disjoint vectors of variables. Let $X$ be a
polytope in the space of $\mathbf{x}$, and $U,$ a polytope in the space of
$\dbinom{\mathbf{w}}{\mathbf{x}}$, with:%
\begin{align}
X  &  :=Conv\left(  \left\{  \left(
\begin{array}
[c]{c}%
8\\
10\\
6
\end{array}
\right)  ,\left(
\begin{array}
[c]{c}%
12\\
15\\
9
\end{array}
\right)  \right\}  \right)  \text{, and}\label{EF_NumEx(a)}\\
U  &  :=\left\{  \dbinom{\mathbf{w}}{\mathbf{x}}\in\mathbb{R}^{4}%
:2\leq\mathbf{0}\cdot\mathbf{x}+\mathbf{w}\leq3\right\}  \label{EF_NumEx(b)}%
\end{align}

We now discuss some key results of Fiorini \textit{et al}. (2015) which are
refuted by $X$ and $U.$

\begin{enumerate}
\item \textit{Refutation of the validity of Definition \ref{EF_Dfn_A1}.}

\begin{enumerate}
\item Note that the following is true for $X$ and $U$:
\begin{equation}
\left(  \mathbf{x}\in X\nLeftrightarrow\left(  \exists\mathbf{w\in}%
\mathbb{R}:\dbinom{\mathbf{w}}{\mathbf{x}}\in U\right)  \right)
.\label{EF_NuimEx(d)}%
\end{equation}
For example,
\begin{equation}
\left(  \exists\mathbf{w\in}\mathbb{R}:\left(
\begin{array}
[c]{c}%
w\\
22.5\\
-50\\
100
\end{array}
\right)  \in U\right)  \nRightarrow\left(  \left(
\begin{array}
[c]{c}%
22.5\\
-50\\
100
\end{array}
\right)  \in X\right)  .\label{EF_NumEx(e)}%
\end{equation}
\ \ Hence, $U$ \textbf{is not} an \textit{extended formulation} of $X$
according to Definition \ref{EF_Dfn_A2}.

\item Observe that the following is also true for $X$ and $U$:
\begin{equation}
X=\left\{  \mathbf{x\in}\mathbb{R}^{3}:\left(  \mathbf{x}=A\cdot
\dbinom{\mathbf{w}}{\mathbf{x}}\mathbf{,}\text{ }\dbinom{\mathbf{w}%
}{\mathbf{x}}\in U\right)  \right\}  \mathbf{,}\text{ where }A\mathbf{=}%
\left[
\begin{array}
[c]{cccc}%
4 & 0 & 0 & 0\\
5 & 0 & 0 & 0\\
3 & 0 & 0 & 0
\end{array}
\right]  . \label{EF_NumEx(f)}%
\end{equation}
In other words, $X$ is the image of $U$ under the linear map $A$. Hence, $U$
\textbf{is} an \textit{extended formulation} of $X$ according to Definition
\ref{EF_Dfn_A1}.

\item It follows from (a) and (b) above, that Definitions \ref{EF_Dfn_A1} and
\ref{EF_Dfn_A2} are in contradiction of each other with respect to $X$ and
$U$. Hence $X$ and $U$ are a refutation of the validity of Definition
\ref{EF_Dfn_A1} (since it is easy to verify the equivalence of Definition
\ref{EF_Dfn_A2} to Definition \ref{EF_Dfn_Std}, which is the ``standard'' definition).
\end{enumerate}

\item \textit{Refutation of ``Theorem 3'' (p.17:10) of Fiorini et al. (2015)}.

The proof of the theorem (``Theorem 3'') hinges on Definition \ref{EF_Dfn_A2}.
The specific statement in Fiorini \textit{et al}. (2015; p. 17:10, lines
26-28) is:%
\begin{align}
&  \text{``}...\text{\textit{Because} }\nonumber\\[0.06in]
&  \mathit{Ax\leq b\Longleftrightarrow\exists y:E}^{=}\mathit{x+F}%
^{=}\mathit{y=g}^{=}\mathit{,}\text{ }\mathit{E}^{\leq}\mathit{x+F}^{\leq
}\mathit{y=g}^{=}\mathit{,}\label{EF_NumEx(g)}\\
&  \text{\textit{each inequality in} }\mathit{Ax\leq b}\text{ \textit{is valid
for all points of} }\mathit{Q}\text{. ...''}\nonumber
\end{align}

The equivalent of (\ref{EF_NumEx(g)}) in terms of $X$ and $U$ is:
\begin{equation}
\mathbf{x}\in X\Longleftrightarrow\exists\mathbf{w\in}\mathbb{R}%
:\dbinom{\mathbf{w}}{\mathbf{x}}\in U. \label{EF_NumEx(h)}%
\end{equation}
Clearly, (\ref{EF_NumEx(h)}) is \textbf{not true}, as we have illustrated in
Part ($1.a$) above. Hence, the proof of ``Theorem 3'' (and therefore,
``Theorem 3'') of Fiorini \textit{et al}. (2015) is refuted by $X$ and $U$.

\item \textit{Refutation of ``Lemma 9'' (p. 17:13-17:14) of Fiorini et al. (2015).}

The first part of the lemma is stated (in Fiorini \textit{et al}. (2015))
thus:%
\begin{align*}
&  \text{``\textit{Lemma 9. Let }P\textit{, }Q\textit{, and }F\textit{\ be
polytopes. Then, the following hold:}}\\
&  \text{\textit{(i) if F is an extension of P, then }xc(F)}\geq
\text{xc(P);\ldots''\ }%
\end{align*}

The proof of this is stated as follows:%
\[
\text{``\textit{Proof. The first part is obvious because every extension of
F\ is in particular an extension of P. }\ldots''}%
\]

The notation ``$xc(\cdot)"$ stands for ``\textit{extension complexity} of
($\cdot$),'' which is defined as (p. 17:9, lines 24-25 of Fiorini \textit{et
al}. (2015)):%
\[
\text{``...\textit{the extension complexity of P is the minimum size (i.e.,
the number of inequalities) of an EF of P}.''}%
\]

The refutation of these for $X$ (as shown in (\ref{EF_NumEx(a)}) above) and
$U$ (as shown in (\ref{EF_NumEx(b)}) above) is as follows.

As shown in Part ($1$) above, $U$ is an \textit{extension} of $X$ according to
Definition \ref{EF_Dfn_A1} (which is central in Fiorini \textit{et al}.
(2015)). This means that $U$ is an \textit{extended formulation} of every one
of the infinitely-many possible $\mathcal{H}$-descriptions of $X$. This would
be true in particular for the $\mathcal{H}$-description below for $X$:
\begin{equation}
X:=\left\{
\begin{array}
[c]{l}%
\mathbf{x\in}\mathbb{R}^{3}:\\
\\
-5x_{1}+4x_{2}\leq0;\\
\text{ }\\
3x_{2}-5x_{3}=0;\text{ }\\
\\
3x_{1}-4x_{3}\leq0;\\
\\
8\leq x_{1}\leq12;\text{ }\\
\\
10\leq x_{2}\leq15;\\
\text{ }\\
6\leq x_{3}\leq9
\end{array}
\right\}  . \label{Counter_Example_Description_2}%
\end{equation}
Clearly, however, we have that:
\begin{equation}
xc(U)\ngeq xc(X). \label{EF_NumEx(i)}%
\end{equation}
Hence, $X$ and $U$ are a refutation of ``Lemma 9'' of Fiorini \textit{et al}.
(2015), being that $U$ is the \textit{extension}, and\textit{\ }$X$, the
\textit{projection,} according to definitions used in Fiorini \textit{et al}. (2015).
\end{enumerate}

\noindent$\square\medskip$
\end{example}

According to Fiorini \textit{et al}. (2015; p. 17:7, Section 1.4, first
sentence; p. 17:11, lines 6-11; p. 17:14, lines 5-6; p.17:16, lines 13-14
after the ``Fig. 4'' caption), their ``Theorem 3'' and ``Lemma 9'' play
pivotal, foundational roles in the rest of their developments. Note that
``Lemma 9'' (of Fiorini \textit{et al.} (2015)) does not depend on any one of
the \textit{extended formulations} definitions used in Fiorini \textit{et al.}
(2015) in particular. Hence, we believe the numerical illustration we have
provided above represents a simple-yet-complete refutation of their
developments when polytopes are described in terms of disjoint sets of
variables. In other words, our counter-example shows that the Fiorini
\textit{et al}. (2015) developments may be valid for models which require (in
a non-redundant way) the natural variables used to describe the standard
polytopes of the problems they consider only.

As can be seen from the illustrative counter-example above, the existence of a
linear map stipulated in Definition \ref{EF_Dfn_A1} is not sufficient to imply
that there exists an \textit{extended formulations} relationship between the
models from which valid/meaningful inferences can be made. We will provide,
below, some insights into the correct meaning/consequence of the existence of
a linear map stipulated in Definition \ref{EF_Dfn_A1}, establishing a
one-to-one correspondence between two models that are stated in disjoint
variable spaces.

\section{Meaning of the existence of a linear transformation}

\noindent We will focus on the more general case (than that of the linear map
stipulated in Definition \ref{EF_Dfn_A1}) of the existence of an affine
transformation which was brought to our attention in private e-mails by
Yannakakis (2013). In the case of polytopes stated in disjoint variable
spaces, if the constraints expressing the affine transformation are
\textit{redundant} for each of the models/polytopes, the implication is that
one model can be used in an ``auxiliary'' way, in order to solve the
optimization problem over the other model, without any reference to/knowledge
of the $\mathcal{H}$-description of that other model. This is shown in Remark
\ref{EF_Insight_Rmk1} below. \ \ \ \ 

\begin{remark}
\label{EF_Insight_Rmk1} \ \noindent

\begin{itemize}
\item Let:

\begin{itemize}
\item $x\in\mathbb{R}^{p}$ and $y\in\mathbb{R}^{q}$ be disjoint vectors of variables;

\item $X:=\{x\in\mathbb{R}^{p}:Ax\leq a\};$

\item $L:=\left\{  \dbinom{x}{y}\in\mathbb{R}^{p+q}:Bx+Cy=b\right\}  $;

\item $Y:=\{y\in\mathbb{R}^{q}:Dy\leq d\};$
\end{itemize}

\noindent(Where: $\ A\in\mathbb{R}^{k\times p};$ $a\in\mathbb{R}^{k};$
$B\in\mathbb{R}^{m\times p};$ $C\in\mathbb{R}^{m\times q};$ $b\in
\mathbb{R}^{m};$ $D\in\mathbb{R}^{l\times q},$ $d\in\mathbb{R}^{l}%
$)$.\smallskip$

\item If $B^{T}B$ is nonsingular, then $L$ can be re-written in the form:%
\begin{align}
&  L=\left\{  \dbinom{x}{y}\in\mathbb{R}^{p+q}:x=\overline{C}y+\overline
{b}\right\}  .\text{ }\nonumber\\[0.06in]
&  \text{(Where: }\overline{C}:=-(B^{T}B)^{-1}B^{T}C\text{, and }\overline
{b}:=(B^{T}B)^{-1}B^{T}b). \label{L_in_Rmk}%
\end{align}
Hence, the linear map stipulated in Definition \ref{EF_Dfn_A1} is simply a
special case of $L$ in which $b=0$ and $B^{T}B$ is nonsingular.

\item Assume that:

\begin{itemize}
\item $L\neq\varnothing$ exists, with constraints that are \textit{redundant}
for $X$ and $Y$, respectively;

\item the non-negativity requirements for $x$ and $y$ are included in the
constraint sets of $X$ and $Y$, respectively; and that:

\item $B^{T}B$ is nonsingular.
\end{itemize}

(This is equivalent to assuming that the more general (affine map) version of
the linear map stipulated in Definition \ref{EF_Dfn_A1} exists.)\smallskip

\item Then, the optimization problem:\smallskip

\textit{Problem LP}$_{0}$:\medskip\newline
\begin{tabular}
[c]{l}%
\ \ \
\end{tabular}
$\left|
\begin{tabular}
[c]{ll}%
$\text{Minimize:}$ & $\alpha^{T}x$\\
& \\
$\text{Subject To:}$ & $x\in X$\\
& \\
\multicolumn{2}{l}{(where $\alpha\in\mathbb{R}^{p}).$}%
\end{tabular}
\ \ \text{ \ }\right.  \medskip$\newline is equivalent to:$\medskip$%
\newline \textit{Problem LP}$_{1}$:\medskip\newline
\begin{tabular}
[c]{l}%
\ \ \
\end{tabular}
$\left|
\begin{tabular}
[c]{ll}%
$\text{Minimize:}$ & $\alpha^{T}x$\\
& \\
$\text{Subject To:}$ & $\dbinom{x}{y}\in L;$ $\ x\in X;$ \ $y\in Y$\\
& \\
\multicolumn{2}{l}{(where $\alpha\in\mathbb{R}^{p}).$}%
\end{tabular}
\ \ \text{ \ }\right.  \bigskip$\newline which is equivalent to the smaller
problem:\medskip\newline \textit{Problem LP}$_{2}$:\medskip\newline
\begin{tabular}
[c]{l}%
\ \ \
\end{tabular}
$\left|
\begin{tabular}
[c]{ll}%
$\text{Minimize:}$ & $\left(  \alpha^{T}\overline{C}\right)  y+\alpha
^{T}\overline{b}$\\
& \\
$\text{Subject To:}$ & $y\in Y$\\
& \\
\multicolumn{2}{l}{(where $\alpha\in\mathbb{R}^{p}).$}%
\end{tabular}
\ \ \text{ \ }\right.  \medskip\medskip$

\item Hence, if $L$ is the graph of a one-to-one correspondence between the
points of $X$ and the points of $Y$ (see Beachy and Blair (2006, pp. 47-59)),
then, the optimization of any linear function of $x$ over $X$ can be done by
first using \textit{Problem LP}$_{\mathit{2}}$ in order to get an optimal $y,$
and then using Graph $L$ to ``retrieve'' the corresponding $x$. Note that the
second term of the objective function of \textit{Problem LP}$_{\mathit{2}}$
can be ignored in the optimization process of \textit{Problem LP}%
$_{\mathit{2}},$ since that term is a constant.\medskip

Hence, if $L$ is derived from knowledge of the $\mathcal{V}$-representation of
$X$ only, then this would mean that the $\mathcal{H}$-representation of $X$ is
not involved in the ``two-step'' solution process (of using \textit{Problem
LP}$_{\mathit{2}}$ and then Graph $L$), but rather, that only the
$\mathcal{V}$-representation of $X$ is involved.\ \ \ 
\end{itemize}

\noindent$\square$
\end{remark}

\section{Conclusions}

We have shown the non-validities of an alternate definition of
\textit{extended formulations} (Definition \ref{EF_Dfn_A1}) which is used in
the Fiorini \textit{et al}. (2011; 2012; 2015) developments when the models
being studied are (or can be) stated in terms of disjoint sets of variables,
and of key results of theirs (specifically, ``Theorem 3'' and ``Lemma 9'' in
Fiorini \textit{et al}. (2015)) which are the foundations for all their other
developments. Hence, the claims in Fiorini \textit{et al}. (2015) that:

\begin{quotation}
``\textit{We solve this question by proving a super-polynomial bound on the
number of }\newline \textit{inequalities in every LP for the TSP.}'' (Fiorini
\textit{et al}. (2015, p.17:2, lines 9-10.)); \medskip

``\textit{We} \textit{also} \textit{prove such unconditional super-polynomial
bounds for the maximum cut }\newline \textit{and} \textit{the}
\textit{maximum} \textit{stable} \textit{set} \textit{problems.''} (Fiorini
\textit{et al}. (2015, p.17:2, lines 10-12.));
\end{quotation}

and

\begin{quotation}
``...\textit{it is} \textit{impossible to prove} \textit{P = NP by means}
\textit{of a polynomial-sized LP that }\newline \textit{e\textit{x}presses any
of these problems.}'' (Fiorini \textit{et al}. (2015, p.17:2, lines 12-13.))
\end{quotation}

\noindent are not supported by the developments in Fiorini \textit{et al}.
(2015), since, as we have shown in this short note, those developments are not
valid for (and are, therefore, not applicable to) models that do not need the
natural/standard variables that are used to describe the standard polytopes
for the problems considered in Fiorini \textit{et al}. (2015). Hence, these
claims are overscoped/overreaching, and refuted by the developments in this
short note. \pagebreak

\end{document}